\documentclass[12pt]{article}

\usepackage{graphics}
\usepackage[percent]{overpic}

\begin{document}

\title{Birefringence of silica hydrogels prepared under high magnetic fields reinvestigated}
\maketitle
\renewcommand{\thefootnote}{\fnsymbol{footnote}}
\noindent
\author{Atsushi Mori\footnote{Corresponding author; phone:~+81-88-656-9417, fax:~+81-88-656-9435, e-mail:~atsushimori@tokushima-u.ac.jp}$^a$,
\renewcommand{\thefootnote}{\arabic{footnote}}
\setcounter{footnote}{0}
Takamasa Kaito\footnote{Present address: KRI Inc., 5-11-151 Torishima, Konohana-ku, Osaka 554-0051, Japan}$^b$,
Hidemitsu Furukawa$^c$,
Masafumi Yamato$^d$,
Kohki Takahashi$^e$} \\[2ex]

{\small\noindent
$^a$Department of Advanced Materials, Institute of Technology and Science, The University of Tokushima, Tokushima 770-8506, Japan \\[2ex]
$^b$Department of Optical System Engineering, Graduate School of Engineering, The University of Tokushima, Tokushima 770-8506, Japan \\[2ex]
$^c$Department of Mechanical Systems Engineering, Yamagata University, Yonezawa, Yamagata 992-8501, Japan \\[2ex]
$^d$Faculty of Urban Environmental Science, Tokyo Metropolitan University, Tokyo 192-0397, Japan \\[2ex]
$^e$High Field Laboratory for Superconducting Materials, Institute for Materials Research, Tohoku University, Sendai 980-8577, Japan \\[2ex]}
\date{Submitted: August 5, 2014; revised: September 8, 2014}
\newpage
\begin{abstract}
Birefringence is an indicator of structural anisotropy of materials.
We measured the birefringence of Pb(II)-doped silica hydrogels prepared under a high magnetic field of various strengths.
Because the silica is diamagnetic, one does not expect the structural anisotropy induced by a magnetic field.
In previous work [Mori A, Kaito T, Furukawa H 2008 \textit{Mater. Lett.} \textbf{62} 3459-3461], we prepared samples in cylindrical cells made of borosilicate glass and obtained a preliminary result indicating a negative birefringence for samples prepared at 5T with the direction of the magnetic field being the optic axis.
We have measured the birefringence of Pn(II)-doped silica hydrogels prepared in square cross-sectional cells made of quartz and reverted the previous conclusion.
Interestingly, the magnetic-influenced silica hydrogels measured have been classified into four classes: two positive birefringent ones, no birefringent one, and negative birefringent one.
Proportionality between birefringence and the strength of magnetic field is seen for the former two. \\[2ex]
\end{abstract}
\newpage
%\sloppy

\section{Introduction}
\subsection{\label{sec:general}General}
Birefringence is a measure of structural anisotropy of materials.
For example, the measure of birefringence is proportional to the nematic order parameter in nematic liquid crystals \cite{Strobl}.
Form and flow birefringence reflects the structural anisotropy of polymers \cite{Doi}.
Stress of transparent materials can be evaluated through measurement of stress birefringence \cite{Born,Hect}.

Birefringence itself can be a function of materials.
Polarization state of light is transformed through birefringent media \cite{Born,Hect}.
In particular, a piece of a birefringent material with a definite retardation, such as a quarter wave plate and a half wave plate, plays a special role in optical systems.
That is, transformation between linear polarization and circular one is carried out through a quarter wave plate and that between right- and left-circular polarizations is carried out through a half wave plate.
To stabilize the value of retardation, solid materials are suitable.

On the other hand, birefringent soft materials have a potential for functional materials owing to their flexibility.
One of characteristics of the soft materials is their stimuli-sensitivity.
For example, nematic order induced by an external field such as an electric field can be controlled by the strength of field.
One can change the direction of fast/slow axis as well as the magnitude of retardation without exchanging the piece relying on this property.
Liquid-crystal variable retarders were now widely used such as in Refs.~\cite{Bueno2000,Heredero2007,Vargas2010,Terrier2010,Wozniak2011,Uribe-Patarroyo2012,Zhang2013,Zhang2014}.
The variable focal liquid-crystal lens \cite{Sato1967} is another example of applications, where the average birefringence is controlled through the variable nematic order induced by an external field.
In turn, we shift our focus to gels.
If one successfully prepared birefringent gels with a definite birefringence $\Delta n$, because the gels can be cut in arbitrary size (say, $d$) easily, one can make a piece of a desired retardation $\mit\Gamma$ $=$ $\Delta n d$.
Of course, there is a tradeoff between the shape stability and the flexibility.

Effect of a magnetic field applied during the gelation on the network structure of organic polymer gels was studied.
Alignment perpendicular to the magnetic field occurred for poly($N$-isopropylacrylamid) gels \cite{Otsuki2006} and agarose gels \cite{Yamamoto2006,Yamamoto2008}.
In cases side-chain group prefers the parallel alignment, polymer chains align perpendicularly to the magnetic field because the side-chain group is basically normal to the main chain.
On the other hand, in cases a group has magnetic moment, a part including such a group prefers orientation in parallel to the magnetic field \cite{Shigekura2005}.
One can intuitively understand the mechanism of magnetic alignment for such polymer gels.

We have, so for, focused on the magnetic alignment of silica hydrogels.
In particular, we studied Pb(II)-doped silica hydrogels prepared in a magnetic field.
Opposed to the organic polymer gels, because silica as well as Pb(II) is diamagnetic, one cannot expect magnetic alignment of silica hydrogels.
However, as described in Sec.~\ref{sec:background}, surprisingly, the magnetic alignment of silica hydrogels did occur.
As a possibility, there is an entirely new mechanism of the magnetic alignment, which may open a new field of science.
Of source, after revealing the new mechanism, new techniques to make alignment of materials whose magnetic alignment has not, so for, been expected will be developed.
One of advantages of the magnetic alignment is that the operation is simple as compared to a programed processing such as a thermal treatment.
Anisotropic gels themselves must possess anisotropic transportation of materials in them.
Such materials can be used as new media in gel permeation chromatography.
If anisotropic silica hydrogels can be solidified, \textit{e.g.}, by drying and then baking, keeping the anisotropy, improvement of mechanical strength in certain direction is expected.
Anisotropically dried silica gels can be used as new media for separation such as column chromatography.
As suggested by a result of previous work \cite{Kaito2006JCG289}, oriented-nanocrystallite dispersed silica materials can be expected.
Such materials can function such as nanophotonic and quantum confinement materials.

In this paper, we reinvestigate birefringence of Pb(II)-doped silica hydrogels prepared in magnetic fields.
A previous conclusion of negative birefringence \cite{Mori2008} will be reverted.
This incorrect conclusion might be due to the use of borosilicate glass cells.
In Sec.~\ref{sec:renew}, renewed results of birefringence measurement for borosilicate glass cells are presented to supplement Sec.~\ref{sec:background}.
The main purpose is to present results of birefringence of samples prepared in quartz cells.

\subsection{\label{sec:background}Background}
In 2006, we grew lead (II) bromide (PbBr${}_2$) crystals using silica hydrogels as a medium of a crystal growth \cite{Kaito2006JCG289,Kaito2006JCG294}.
The silica hydrogels were made from sodium metasilicate (Na$_2$SiO$_3\cdot$9H$_2$O) aqueous solution by adding a concentrated acetic acid solution.
Na$_2$SiO$_3$-based silica hydrogels were often used as media of crystal growths \cite{Panadita2001,Kusumoto2005,Kumar2007MRB,Kumar2007ML,Yanagiya2007}.
As a source of Pb(II) we added a lead (II) nitrite solution.
This mixed solution was settled in a magnetic fields of $B$ $=$ 5T to prepare the silica hydrogels.
We observed an ordered array of PbBr${}_2$ nanocrystallites with their crystallographic axes aligned to the direction of magnetic field as a result of the crystal growth.
It is suggested from the fact that no magnetic field was applied during the crystal growth that a structural anisotropy was formed in the silica hydrogels.
To detect the anisotropy in the silica gels, we \cite{Mori2008} measured birefringence $\Delta n$ and performed scanning microscopic light scattering \cite{Furukawa2003} (SMILS).
Though improvement of equipment of the S\`{e}narmont measurement is now in progress (thus, the values will be revised), at that time we detected a negative birefringence of magnitude on the order of $10^{-6}$ with the direction of the magnetic fields being as the optic axis.
In addition, those values were affected certainly by the sample cells of borosilicate glass, as described in the next paragraph concerning our successive study \cite{Tomita2010}.
However, the results of SMILS did certainly indicate ordering; the width of the distribution of relaxation time of autocorrelation function of the electric fields of scattered light was smaller for samples exhibiting birefringence than for ones exhibiting no birefringence.
Even if one suspects the results of birefringence, we can say that the samples are classified into two, one has a narrow size distribution of the gel network and the other a wide one.

\begin{figure}[htb]
\begin{center}
\includegraphics[width=0.45\textwidth, keepaspectratio, clip]{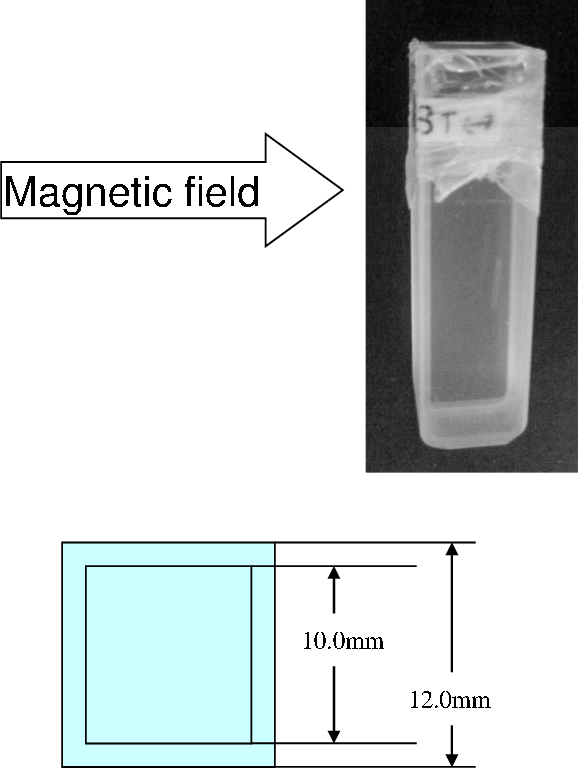}
\caption{\label{fig:cell}A photograph and cross-sectional dimension of cells used in Ref.~\cite{Tomita2010} and the present study.}
\end{center}
\end{figure}

\begin{figure}[htb]
\begin{center}
\includegraphics[width=0.45\textwidth, keepaspectratio, clip]{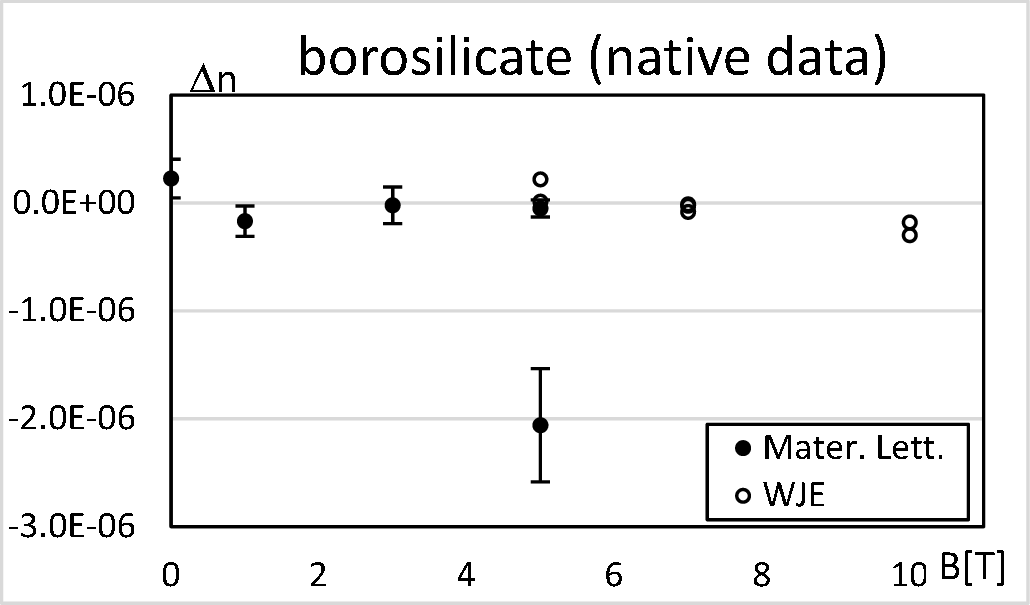}
\caption{\label{fig:boronative}Date of Fig.~3 of Ref.~\cite{Mori2008} (closed circles) and those of Fig.~1 of Ref.~\cite{Tomita2010} (open circles).}
\end{center}
\end{figure}

We extended birefringence measurement of Pb(II)-doped silica hydorgels, which were prepared in magnetic fields up to 10T \cite{Tomita2010}.
As sample cells, square cross-sectional cells of both of borosilicate glass and quartz were used.
A photograph of a cell is shown in Fig.~\ref{fig:cell} with a illustration of the cross section.
We note that a special care should be paid to input the light to the center of the cell for the case of cylindrical cells.
Or, inaccuracy will be accompanied.
This is a geometrical situation, independent of the cell materials.
The method of birefringence measurement was the same as in a previous study \cite{Mori2008}.
In this paragraph, we pick up the results of birefringence measurement of samples prepared in borosilicate glass cells in order to clarify the reason why we were conducted to an incorrect conclusion of the negative birefringence.
The results, i.e., native data, of Refs.~\cite{Mori2008,Tomita2010} are plotted together in Fig.~\ref{fig:boronative}.
Date of Fig.~3 of Ref.~\cite{Mori2008} (closed circles) and those of Fig.~1 of Ref.~\cite{Tomita2010} (open circles) are plotted together.
For the former, we have drawn the standard deviation as the full length of error bars, while the half length was the standard deviation in Ref.~\cite{Mori2008}.
The scale of vertical axis is the same as in Ref.~\cite{Mori2008}, and thus different from that in Ref.~\cite{Tomita2010}.
Nevertheless, one can still confirm a negative slope for the open circles.
A negative slope for data of Ref.~\cite{Tomita2010} is, nevertheless, still seen.
We have added data for borosilicate glass cells (Sec.~\ref{sec:renew}), but this trend remains.
In Ref.~\cite{Tomita2010}, the slope was magnified, and thus we harried a conclusion that a decreasing property of $\Delta n(B)$ and the negative birefringence were reproduced.

In previous work \cite{Tomita2010}, as mentioned just above, we insisted the negative birefringence and straggled to explain the results indicating the negative birefringence for borosilicate glass cells as the effect of cell wall surface.
However, if the surface effect is involved, the behaviors for both sample cells at hight magnetic fields, where the field effect overcomes the surface effect, should exhibit the same trend.
That is, the intercept on $B$ $=$ 0 is merely shifted due to the surface effect and the asymptotic behavior for high magnetic field should be the same.
The slopes of $\Delta n (B)$'s for borosilicate glass cells and quartz cells were entirely different from each other. 
In this respect, we infer that unexpected effects are involved in samples prepared in the borosilicate glass cells.
Borosilicate glass cells are not suitable for birefringence measurement.

Before proceeding to the next section, we would like to give a note on the stimuli-sensitivity of soft matters.
In Sec.~\ref{sec:general} we have introduced the variable retarders and variable focal lens.
Those are examples of the stimuli-sensitivity after preparation of materials.
On the other hand, the present concern is on the stimuli-sensitivity during materials preparation.
In this respect, stability of the structure constructed in a magnetic field may be worried.
This is, the magnetically aligned state is not the true equilibrium one.
We already confirmed that the appearance of the gel samples was not changed for years if the vessels were tightly sealed and the same measured birefringence property remained.
It means that evaporation of solvent degrades the samples.
To test the thermal stability, one should raise the temperature without evaporation of the solvent.

\section{Materials and Methods}
\subsection{Materials}
At first, 50g Na${}_2$SiO${}_3\cdot$H${}_2$O was resolved in 103ml distilled water as described in Ref.~\cite{Kaito2006JCG289}; and then 8ml of this aqueous solution was diluted with 8 ml distilled water.
16 ml concentrated acetic acid aqueous solution was added with stirring and then 6.4ml of 1M lead (II) nitrate aqueous solution was added.
This solution was stirred for 2 hours.
And then, the mixture was separated into test tubes (a photograph and an illustration of cross section are shown in Fig~\ref{fig:cell}) and sealed.
After that, they were put in equipment of a high magnetic field, a superconducting magnet (JMTD-10100M, Japan Magnet Technologies Co.).
Five different magnetic fields (3, 5, 7, 9, and 10T) were applied by setting the samples at different positions.
The samples were gellated by settlement for a week.
All materials treatments were made at room temperature (298K).

\subsection{\label{sec:biref}Birefringence measurement}
Intensity $I(\theta)$ of light emerging from the analyze in the S\`{e}narmont method was measured to determine the extinction angle.
A spectrometer (Ocean Optics, USB4000) was mounted on a conventional polarization microscope (Nikon OPTIPHOT2-POL) to obtain the intensity $I$ as a function of the rotation angle $\theta$ of the analyzer.
By virtue of obtaining the function $I(\theta)$, instead of detection of the extinction angle by naked eyes as done in a usual S\`{e}narmont method using options to a conventional polarization microscope, we could detect small extinction angles such as less than a degree.
If no sample, the transmitted light extincts for the configuration of the crossed polarizers.
If a birefringent sample is inserted with the axis pointing 45$^\circ$ with respect to the transmitting axis of the polarizer, the extinction angle changes.
We fitted the intensity by $I$ $=$ $A \cos(2\theta -\delta)$ $+$ $C$ to determine the extinction angle $\delta$[rad] $=$ $2\pi\mit\Gamma/\lambda$, where $\lambda$ is the wavelength of light source, commonly the mercury light, whose $\lambda$ being 546nm.
In the present study, the mercury light was employed.
All measurements were done at room temperature (298K).
Photons were counted per 100ms for a few minutes to obtain a time-averaged intensity $I$ at one $\theta$, and the standard deviation was also calculated.
For fitting of $I(\theta)$, nineteen measurements at every $10^\circ$ were done for one point in a sample.
Therefore, it took one hour or more to obtain $I(\theta)$'s for one point in a sample.
This method was so time-consuming; thus, the number of measured data was limited.
The measure of birefringence, $\Delta n$, is obtained from $\mit\Gamma$ = $\Delta n d$, where $d$ is the thickness of the sample.

\section{Results and Discussion}
\begin{figure}[htb]
\begin{center}
\includegraphics[width=0.45\textwidth, keepaspectratio, clip]{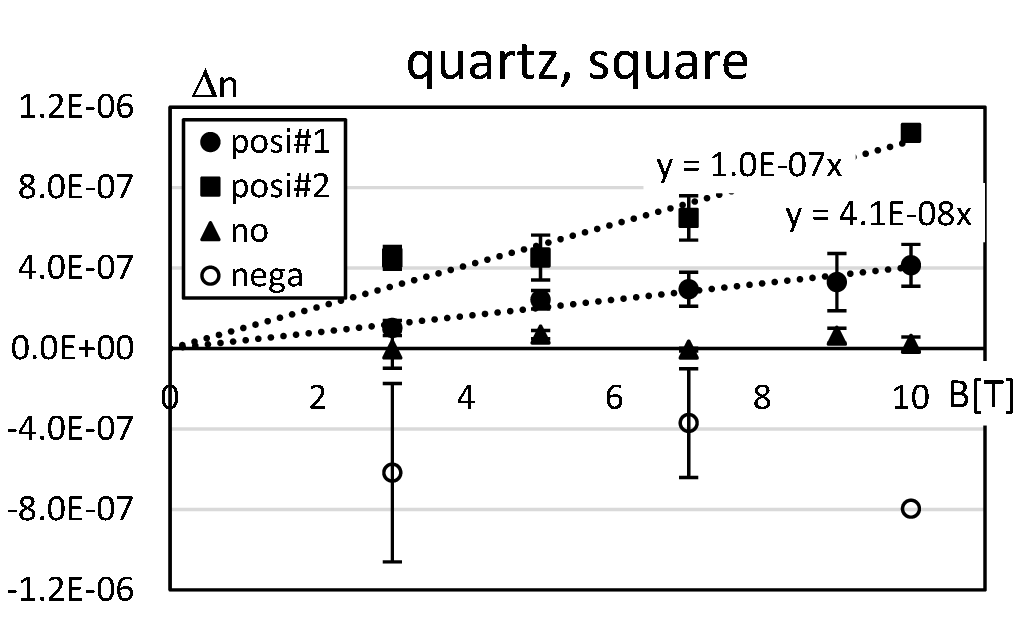}
\caption{\label{fig:quartz}Birefringence of Pb(II)-doped silica hydrogels prepared in quartz cells.}
\end{center}
\end{figure}

As mentioned in previous work \cite{Tomita2010}, we selected cells whose optical path difference (retardation) $\mit\Gamma$ without sample was on the order of $10^{-1}$nm or less (see, for details of $\mit\Gamma$, Sec.~\ref{sec:biref}).
Results obtained according to the method described in Sec.~\ref{sec:biref} are listed in Table~\ref{tab:quartz}.
At first, averages over individual samples were taken.
We found that those results are classified into four groups (two groups of positive birefringence, no birefringence one, and negative birefringence one).
We took averages over each groups.
In sample average, data belonging to different groups were treated as different ones.
We included data in previous work \cite{Tomita2010}.
We plots those results in Fig~\ref{fig:quartz}.

\begin{table*}[htbp]
\begin{center}
\caption{\label{tab:quartz}Date for quartz cells, newly measured in this study.}
\begingroup
\setlength{\tabcolsep}{4.4pt}
\begin{tabular*}{\textwidth}{c|cccccccccccc} \hline
\multicolumn{1}{c|}{\small $B$[T]} & \multicolumn{12}{c}{10} \\ \hline
\multicolumn{1}{c|}{\tiny sample} & \multicolumn{1}{c|}{\tiny 10Q1} & \multicolumn{2}{c|}{\tiny 10Q2} & \multicolumn{3}{c|}{\tiny 10Q3} & \multicolumn{3}{c|}{\tiny 10Q4} & \multicolumn{3}{c}{\tiny 10Q5} \\ \hline
\multicolumn{1}{c|}{\tiny position} & \multicolumn{1}{c|}{1} & \multicolumn{1}{c}{1} & \multicolumn{1}{c|}{2} & \multicolumn{1}{c}{1} & \multicolumn{1}{c}{2} & \multicolumn{1}{c|}{3} & \multicolumn{1}{c}{1} & \multicolumn{1}{c}{2} & \multicolumn{1}{c|}{3} & \multicolumn{1}{c}{1} & \multicolumn{1}{c}{2} & \multicolumn{1}{c}{3}  \\ \hline
\multicolumn{1}{c|}{\tiny $10^7\Delta n$} & \multicolumn{1}{c|}{\tiny 3.84} & \multicolumn{1}{c}{\tiny 0.284} & \multicolumn{1}{c|}{\tiny 0.714} & \multicolumn{1}{c}{\tiny 6.08} & \multicolumn{1}{c}{\tiny 4.79} & \multicolumn{1}{c|}{\tiny 3.39} & \multicolumn{1}{c}{\tiny 6.74} & \multicolumn{1}{c}{\tiny 4.45} & \multicolumn{1}{c|}{\tiny 2.59} & \multicolumn{1}{c}{\tiny 1.07} & \multicolumn{1}{c}{\tiny 0.840} & \multicolumn{1}{c}{\tiny 0.922} \\ \hline
\hline
\multicolumn{1}{c|}{\small $B$[T]} & \multicolumn{3}{c|}{10} & \multicolumn{9}{c}{9} \\ \hline
\multicolumn{1}{c|}{\tiny sample} & \multicolumn{1}{c|}{\tiny 10Q6} & \multicolumn{1}{c|}{\tiny 10Q7} & \multicolumn{1}{c|}{\tiny 10Q8}& \multicolumn{3}{c|}{\tiny 9Q1} & \multicolumn{3}{c|}{\tiny 9Q2} & \multicolumn{3}{c}{\tiny 9Q3} \\ \hline
\multicolumn{1}{c|}{\tiny position} & \multicolumn{1}{c|}{1} & \multicolumn{1}{c|}{1} & \multicolumn{1}{c|}{1} & \multicolumn{1}{c}{1} & \multicolumn{1}{c}{2} & \multicolumn{1}{c|}{3} & \multicolumn{1}{c}{1} & \multicolumn{1}{c}{2} & \multicolumn{1}{c|}{3} & \multicolumn{1}{c}{1} & \multicolumn{1}{c}{2} & \multicolumn{1}{c}{3} \\ \hline
\multicolumn{1}{c|}{\tiny $10^7\Delta n$} & \multicolumn{1}{c|}{\tiny $-0.537$} & \multicolumn{1}{c|}{\tiny $-7.96$} & \multicolumn{1}{c|}{\tiny 10.7} & \multicolumn{1}{c}{\tiny 4.92} & \multicolumn{1}{c}{\tiny 2.07} & \multicolumn{1}{c|}{\tiny 0.261} & \multicolumn{1}{c}{\tiny 0.166} & \multicolumn{1}{c}{\tiny 1.52} & \multicolumn{1}{c|}{\tiny 1.37} & \multicolumn{1}{c}{\tiny 5.08} & \multicolumn{1}{c}{\tiny 4.63} & \multicolumn{1}{c}{\tiny 2.82} \\ \hline
\hline
\multicolumn{1}{c|}{\small $B$[T]} & \multicolumn{3}{c|}{9} & \multicolumn{9}{c}{7} \\ \hline
\multicolumn{1}{c|}{\tiny sample} & \multicolumn{3}{c|}{\tiny 9Q4} & \multicolumn{3}{c|}{\tiny 7Q1}& \multicolumn{1}{c|}{\tiny 7Q2} & \multicolumn{2}{c|}{\tiny 7Q3} & \multicolumn{1}{c|}{\tiny 7Q4} & \multicolumn{2}{c}{\tiny 7Q5} \\ \hline
\multicolumn{1}{c|}{\tiny position} & \multicolumn{1}{c}{1} & \multicolumn{1}{c}{2} & \multicolumn{1}{c|}{3} & \multicolumn{1}{c}{1} & \multicolumn{1}{c}{2} & \multicolumn{1}{c|}{3} & \multicolumn{1}{c|}{1} & \multicolumn{1}{c}{1} & \multicolumn{1}{c|}{2} & \multicolumn{1}{c|}{1} & \multicolumn{1}{c}{1} & \multicolumn{1}{c}{2} \\ \hline
\multicolumn{1}{c|}{\tiny $10^7\Delta n$} & \multicolumn{1}{c}{\tiny 1.40} & \multicolumn{1}{c}{\tiny 6.82} & \multicolumn{1}{c|}{\tiny 2.59} & \multicolumn{1}{c}{\tiny 8.53} & \multicolumn{1}{c}{\tiny 6.80} & \multicolumn{1}{c|}{\tiny 0.261} & \multicolumn{1}{c|}{\tiny $-13.3$} & \multicolumn{1}{c}{\tiny $-4.19$} & \multicolumn{1}{c|}{\tiny $-0.949$} & \multicolumn{1}{c|}{\tiny $-0.806$} & \multicolumn{1}{c}{\tiny 3.92} & \multicolumn{1}{c}{\tiny 3.09} \\ \hline
\hline
\multicolumn{1}{c|}{\small $B$[T]} & \multicolumn{3}{c|}{7} & \multicolumn{9}{c}{5} \\ \hline
\multicolumn{1}{c|}{\tiny sample} & \multicolumn{1}{c|}{\tiny 7Q6} & \multicolumn{2}{c|}{\tiny 7Q7} &\multicolumn{3}{c|}{\tiny 5Q1} & \multicolumn{3}{c|}{\tiny 5Q2} & \multicolumn{3}{c}{\tiny 5Q3} \\ \hline
\multicolumn{1}{c|}{\tiny position} & \multicolumn{1}{c|}{1} & \multicolumn{1}{c}{1} & \multicolumn{1}{c|}{2} & \multicolumn{1}{c}{1} & \multicolumn{1}{c}{2} & \multicolumn{1}{c|}{3} & \multicolumn{1}{c}{1} & \multicolumn{1}{c}{2} & \multicolumn{1}{c|}{3} & \multicolumn{1}{c}{1} & \multicolumn{1}{c}{2} & \multicolumn{1}{c}{3}  \\ \hline
\multicolumn{1}{c|}{\tiny $10^7\Delta n$} & \multicolumn{1}{c|}{\tiny 1.11} & \multicolumn{1}{c}{\tiny $-0.174$} & \multicolumn{1}{c|}{\tiny $-0.118$} & \multicolumn{1}{c}{\tiny 6.29} & \multicolumn{1}{c}{\tiny 2.44} & \multicolumn{1}{c|}{\tiny 1.30} & \multicolumn{1}{c}{\tiny 7.97} & \multicolumn{1}{c}{\tiny 5.63} & \multicolumn{1}{c|}{\tiny 3.50} & \multicolumn{1}{c}{\tiny 4.49} & \multicolumn{1}{c}{\tiny 1.63} & \multicolumn{1}{c}{\tiny 0.688} \\ \hline
\hline
\multicolumn{1}{c|}{\small $B$[T]} & \multicolumn{11}{c|}{5} & \multicolumn{1}{c}{-} \\ \hline
\multicolumn{1}{c|}{\tiny sample} & \multicolumn{3}{c|}{\tiny 5Q4} & \multicolumn{1}{c|}{\tiny 5Q5} & \multicolumn{2}{c|}{\tiny 5Q6} & \multicolumn{1}{c|}{\tiny 5Q7} & \multicolumn{1}{c|}{\tiny 5Q8} & \multicolumn{1}{c|}{\tiny 5Q9} & \multicolumn{2}{c|}{\tiny 5Q10} & \multicolumn{1}{c}{-} \\ \hline
\multicolumn{1}{c|}{\tiny position} & \multicolumn{1}{c}{1} & \multicolumn{1}{c}{2} & \multicolumn{1}{c|}{3} & \multicolumn{1}{c|}{1} & \multicolumn{1}{c}{1} & \multicolumn{1}{c|}{2} & \multicolumn{1}{c|}{1} & \multicolumn{1}{c|}{1} & \multicolumn{1}{c|}{1} & \multicolumn{1}{c|}{1} & \multicolumn{1}{c|}{2} & \multicolumn{1}{c}{-} \\ \hline
\multicolumn{1}{c|}{\tiny $10^7\Delta n$} & \multicolumn{1}{c}{\tiny 1.24} & \multicolumn{1}{c}{\tiny $-0.752$} & \multicolumn{1}{c|}{\tiny 0.485} & \multicolumn{1}{c|}{\tiny 1.92} & \multicolumn{1}{c}{\tiny 2.72} & \multicolumn{1}{c|}{\tiny 2.62} & \multicolumn{1}{c|}{\tiny 1.89} & \multicolumn{1}{c|}{\tiny $-0.214$} & \multicolumn{1}{c|}{\tiny 0.845} & \multicolumn{1}{c}{\tiny 0.397} & \multicolumn{1}{c|}{\tiny 0.632} & \multicolumn{1}{c}{-} \\ \hline
\hline
\multicolumn{1}{c|}{\small $B$[T]} & \multicolumn{12}{c}{3} \\ \hline
\multicolumn{1}{c|}{\tiny sample} & \multicolumn{2}{c|}{\tiny 3Q1} & \multicolumn{1}{c|}{\tiny 3Q2} & \multicolumn{3}{c|}{\tiny 3Q3} & \multicolumn{2}{c|}{\tiny 3Q4} & \multicolumn{1}{c|}{\tiny 3Q5} & \multicolumn{3}{c}{\tiny 3Q6} \\ \hline
\multicolumn{1}{c|}{\tiny position} & \multicolumn{1}{c}{1} & \multicolumn{1}{c|}{2} & \multicolumn{1}{c|}{1} & \multicolumn{1}{c}{1} & \multicolumn{1}{c}{2} & \multicolumn{1}{c|}{3} & \multicolumn{1}{c}{1} & \multicolumn{1}{c|}{2} & \multicolumn{1}{c|}{1} & \multicolumn{1}{c}{1} & \multicolumn{1}{c}{2} & \multicolumn{1}{c}{3} \\ \hline
\multicolumn{1}{c|}{\tiny $10^7\Delta n$} & \multicolumn{1}{c}{\tiny $-10.6$} & \multicolumn{1}{c|}{\tiny $-9.76$} & \multicolumn{1}{c|}{\tiny $-1.11$} & \multicolumn{1}{c}{\tiny 0.611 } & \multicolumn{1}{c}{\tiny 0.112} & \multicolumn{1}{c|}{\tiny $-0.135$} & \multicolumn{1}{c}{\tiny $-0.543$} & \multicolumn{1}{c|}{\tiny 0.769} & \multicolumn{1}{c|}{\tiny 1.55} & \multicolumn{1}{c}{\tiny 5.62} & \multicolumn{1}{c}{\tiny 4.53} & \multicolumn{1}{c}{\tiny 3.36} \\ \hline
\hline
\multicolumn{1}{c|}{\small $B$[T]} & \multicolumn{2}{c|}{3} & \multicolumn{10}{c}{-}\\ \hline
\multicolumn{1}{c|}{\tiny sample} & \multicolumn{2}{c|}{\tiny 3Q7} & \multicolumn{10}{c}{-} \\ \hline
\multicolumn{1}{c|}{\tiny position} & \multicolumn{1}{c}{1} & \multicolumn{1}{c|}{2} & \multicolumn{10}{c}{-} \\ \hline
\multicolumn{1}{c|}{\tiny $10^7\Delta n$}  & \multicolumn{1}{c}{\tiny $-9.99$} & \multicolumn{1}{c|}{\tiny $-3.72$} & \multicolumn{10}{c}{-} \\ \hline
\end{tabular*}
\endgroup
\end{center}
\end{table*}

As shown by dotted lines, surprisingly, proportionality appears well for tow positive birefringence groups.
Points are distributed over entire lines for both branches.
Also, no birefringence data are distributed up to 10T.
Data for the negative birefringence group are scattered and accompany large errors.
Thus, there is a possibility that the negative birefringence was not originated in the magnetic alignment.
We wish to exclude those data from further discussion.
There may be three phases in Pb(II)-doped silica hydrogels prepared in magnetic fields.
In general, one phase is stable one and the others are metastable.
Birefringence of Pb(II)-doped silica hydrogels prepared in quartz cells is plotted in Fig.~\ref{fig:quartz}.
As mentioned just above, date are classified into four groups: two positive birefringence ones (^^ ^^ posi\#1" and ^^ ^^ posi\#2"), no birefringence one (^^ ^^ no"), and negative birefringence one (^^ ^^ nega").
In previous work \cite{Mori2008}, we made an image of the first-order phase transition from no birefringence phase to birefringence one as the strength of magnetic field increases.
On the other hand, on the basis of the present results, one can expect a successive transition from no birefringent phase to low-positive birefringence phase and then to high-positive birefringence phase as the strength of magnetic field increases.
In low-$B$ region the stable phase should be the no birefringence one, in middle-$B$ region the stable phase should be the low-positive one, and in high-$B$ region the stable phase should be the high-positive birefringence one.
In usual first-order phase transition, there are coexistence regions and at the edge of the coexistence region the curve of state equation of metastable phase bends.
Along with the surprisingly good proportionality, wideness of the metastable regions are also remarkable.

Existence of two aligned phases in magnetically-aligned Pb(II)-doped silica hydrogels may be a new discovery.
As opposed to the smectic phase, where plenty of subphases exist, nematic phase exhibits no such plenty --- in this respect, the discovery of the transparent nematic phase was striking \cite{Yamamoto2001}.
There were, however, controversies on reproducibility and then it was said that the transparent nematic phase is a metastable phase; controversies still remain such as by Ref.~\cite{Chen2011}.
However, from the fact that data points are distributed over wide $B$ range, two ^^ ^^ nematic" phases in the magnetically-aligned Pb(II)-doped silica hydrogels have enough stability.

\section{Concluding remarks}
We have reverted a conclusion of previous work \cite{Mori2008} on the sign of birefringence of Pb(II)-doped silica hydrogels prepared in magnetic fields.
By preparing samples in quartz cells, we have obtained positive birefringence samples along with no birefringence samples.
Besides the no birefringence samples (phase), we could classified positive birefringence samples into two groups on the basis of the proportional coefficient between the birefringence $\Delta n$ and the strength of magnetic field $B$.
Surprisingly, the proportionality is well for both two phases.
One can expect a successive phase transition among, at least, three phases.
We wish to postpone the accurate determination of the proportional coefficients.
Along with the accurate proportional coefficients, if the phase transition points are determined, one can develop a method to process silica hydrogels possessing desired $\Delta n$ values.

Difference in structures of gel network of those phases will be helpful in characterizing the phase transition.
Difference in two positive birefringence phases are especially of interest.
Ordering of the gel network occurs in the birefringence phases.
Degree of the order, and thus the structure, should be different in these phases.
In applications such separation, controlling the structure is a key.

In previous work \cite{Mori2008}, perpendicular alignment of silica gel network to the magnetic field was speculated.
We will reconsider on the direction of magnetic alignment of Pb(II)-doped silica hydrogels in very near future.
Fractal property was found in silica hydorogels \cite{Schaefer1984,Ferri1991}.
If anisotropy is introduced, self-affine analysis should be done.
This will be one of future studies.
From the small birefringence, the structural anisotropy must be small.
Preparation in a more strong magnetic field will bring the appreciable anisotropy.

To bring a stability against the evaporation of solvent, the solvent of gel can be replaced with non-voltaic one.
This is one of future studies.
Improvement of thermal stability can also be expected.
It will enable a test on effect of thermal treatment.

\section*{Acknowledgment}
The samples were prepared using equipment of the High Field Laboratory for Superconducting Materials, Institute for Materials Research, Tohoku University.
It is noted that permission of reuse of published figures have been obtainnd from the publusher of Ref.~\cite{Tomita2010} (regarding Ref.~\cite{Mori2008}, the authors retain a right to resuse figures published their own paper) although Figs.~\ref{fig:boronative} and \ref{fig:quartz} as well as Fig.~\ref{fig:bororenew} were not exact reuses.

\appendix
\section{\label{sec:renew}Birefringence for borosilicate glass cells renewed}
\begin{figure}[htb]
\begin{center}
\includegraphics[width=0.45\textwidth, keepaspectratio, clip]{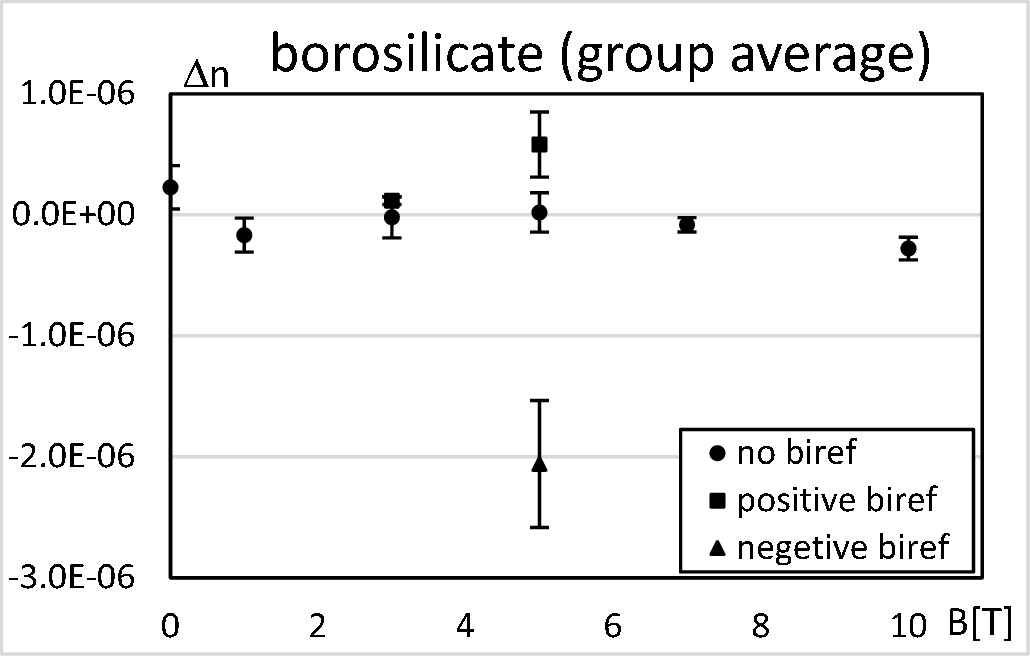}
\caption{\label{fig:bororenew}Renewed results of birefringence for borosilicate glass cells.}
\end{center}
\end{figure}

We have added data for borosilicate glass cells to those shown in Fig~\ref{fig:boronative}.
Added data are listed in Table~\ref{tab:added}.
Samples 5-Ref.\cite{Mori2008} and 3-Ref.\cite{Mori2008} are data which were not shown in Fig.~3 of Ref.~\cite{Mori2008} (out of range).
We were not interested in the positive branch at that time.
Those data are included in the renewed results of Fig.~\ref{fig:bororenew}.
Aiming at drawing the curve of negative branch in high-$B$ region, samples prepared in $B$ $=$ 5, 7, and 10T were investigated.
Unfortunately, we could not obtain results belonging to the negative branch.
Instead, data for the positive branch were obtained for 5T samples (positions 1 and 2 of 5B1 sample and position 1 of 5B2 sample).
As for quartz cells, cells whose $\mit\Gamma$ without sample was on the order of $10^{-1}$ were selected.
To make plots of Fig.~\ref{fig:bororenew}, at first we took averages over individual samples, and then averages were taken for three groups (positive, negative, and no birefringence) -- as mentioned just above, we had no data added for the negative branch.
In sample average, data belonging to different branches were treated as different ones.
Also, data which were out of range of Fig.~3 of Ref.~\cite{Mori2008} were shown in Table~\ref{tab:added} and included in Fig.~\ref{fig:bororenew}.

\begin{table*}[htb]
\begin{center}
\caption{\label{tab:added}Added date for borosilicate glass cells.}
\begingroup
\setlength{\tabcolsep}{3.95pt}
\begin{tabular*}{\textwidth}{c|ccccccccc} \hline
\multicolumn{1}{c|}{$B$[T]} & \multicolumn{9}{c}{10} \\ \hline
\multicolumn{1}{c|}{\tiny sample} & \multicolumn{3}{c|}{10B1} & \multicolumn{3}{c|}{10B2} & \multicolumn{3}{c}{10B3} \\ \hline
\multicolumn{1}{c|}{\tiny position} & \multicolumn{1}{c}{1} & \multicolumn{1}{c}{2} & \multicolumn{1}{c|}{3} & \multicolumn{1}{c}{1} & \multicolumn{1}{c}{2} & \multicolumn{1}{c|}{3} & \multicolumn{1}{c}{1} & \multicolumn{1}{c}{2} & \multicolumn{1}{c}{3} \\ \hline
\multicolumn{1}{c|}{\small $10^7\Delta n$} & \multicolumn{1}{c}{\small $-1.92$} & \multicolumn{1}{c}{\small $-1.40$} & \multicolumn{1}{c|}{\small $-2.08$} & \multicolumn{1}{c}{\small $-5.44$} & \multicolumn{1}{c}{\small $-3.42$} & \multicolumn{1}{c|}{\small $-2.20$} & \multicolumn{1}{c}{\small $-4.45$} & \multicolumn{1}{c}{\small $-3.53$} & \multicolumn{1}{c}{\small $-1.86$} \\ \hline
\hline
\multicolumn{1}{c|}{$B$[T]} & \multicolumn{6}{c|}{7} & \multicolumn{1}{c|}{-} & \multicolumn{2}{c}{5} \\ \hline
\multicolumn{1}{c|}{\tiny sample} & \multicolumn{3}{c|}{7B1} & \multicolumn{3}{c|}{7B2} & \multicolumn{1}{c|}{-} & \multicolumn{2}{c}{5-Ref.\cite{Mori2008}} \\ \hline
\multicolumn{1}{c|}{\tiny position} & \multicolumn{1}{c}{1} & \multicolumn{1}{c}{2} & \multicolumn{1}{c|}{3} & \multicolumn{1}{c}{1} & \multicolumn{1}{c}{2} & \multicolumn{1}{c|}{3} & \multicolumn{1}{c|}{-} & \multicolumn{1}{c|}{1} & \multicolumn{1}{c}{2} \\ \hline
\multicolumn{1}{c|}{\small $10^7\Delta n$} & \multicolumn{1}{c}{\small $-2.55$} & \multicolumn{1}{c}{\small $-0.218$} & \multicolumn{1}{c|}{\small $-0.986$} & \multicolumn{1}{c}{\small $-0.135$} & \multicolumn{1}{c}{\small $-0.491$} & \multicolumn{1}{c|}{\small $-0.402$} & \multicolumn{1}{c|}{-} & \multicolumn{1}{c|}{1.63} & \multicolumn{1}{c}{0.720} \\ \hline
\hline
\multicolumn{1}{c|}{$B$[T]} & \multicolumn{6}{c|}{5} & \multicolumn{1}{c|}{-} & \multicolumn{2}{c}{3}\\ \hline
\multicolumn{1}{c|}{\tiny sample} & \multicolumn{3}{c|}{5B1} & \multicolumn{3}{c|}{5B2} & \multicolumn{1}{c|}{-} & \multicolumn{2}{c}{3-Ref.\cite{Mori2008}}\\ \hline
\multicolumn{1}{c|}{\tiny position} & \multicolumn{1}{c}{1} & \multicolumn{1}{c}{2} & \multicolumn{1}{c|}{3} & \multicolumn{1}{c}{1} & \multicolumn{1}{c}{2} & \multicolumn{1}{c|}{3} & \multicolumn{1}{c|}{-} & \multicolumn{1}{c}{1} & \multicolumn{1}{c}{2} \\ \hline
\multicolumn{1}{c|}{\small $10^7\Delta n$} & \multicolumn{1}{c}{$6.75$} & \multicolumn{1}{c}{$7.11$} & \multicolumn{1}{c|}{$2.28$} & \multicolumn{1}{c}{$8.97$} & \multicolumn{1}{c}{\small $-3.50$} & \multicolumn{1}{c|}{$1.10$} & \multicolumn{1}{c|}{-} & \multicolumn{1}{c}{1.09} & \multicolumn{1}{c}{1.96} \\ \hline
\end{tabular*}
\endgroup
\end{center}
\end{table*}

Fig.~\ref{fig:bororenew} is essentially the same as Fig~\ref{fig:boronative}, except for the positive branch (squares at 3 and 5T).
Thus, we can still insist the negative birefringence.
However, a possibility has arisen that the positive branch is true behavior and the point indicating the negative birefringence (triangle) is an exceptional one.
That is, \textit{e.g.}, deformation after sample preparation may cause unexpected birefringence behavior.
We wish to postpone the conclusion on the sign of birefringence for borosilicate glass cells.

\end{document}